\documentstyle[aps,prl,floats]{revtex}

\newcommand{\nab}{\mbox{\boldmath $\nabla$}}
\newcommand{\nhat}{\hat{\bf n}}
\newcommand{\psihat}{\hat{\Psi}}

\begin{document}

\twocolumn[\hsize\textwidth\columnwidth\hsize\csname @twocolumnfalse\endcsname

\draft

\title{Theory of Quantum Hall Nematics}

\author{Leo Radzihovsky$^1$ and Alan T. Dorsey$^2$}
\address{$^1$Department of Physics, University of Colorado, 
Boulder, Colorado 80309} 
\address{$^2$Department of Physics, University of Florida, Gainesville,
Florida 32611-8440}
\date{\today}

\maketitle

\begin{abstract}
  
  Transport measurements on two dimensional electron systems in
  moderate magnetic fields suggest the existence of a spontaneously
  orientationally-ordered, compressible liquid state.  We develop and
  analyze a microscopic theory of such a ``quantum Hall nematic'' (QHN)
  phase, predict the existence of a novel, highly anisotropic $q^3$
  density-director mode, find that the $T=0$ long-range orientational
  order is unstable to weak disorder, and compute the tunneling into
  such a strongly correlated state. This microscopic approach is
  supported and complemented by a hydrodynamic model of the QHN,
  which, in the dissipationless limit, reproduces the modes of the
  microscopic model.

\end{abstract}

\vspace{-0.2cm}
\pacs{PACS:
        73.40.Hm,       
        73.50.Jt,       
        73.20.Mf,       
        64.70.Md        
}
]

Recent transport measurements on high mobility two dimensional (2d)
electron systems \cite{lilly99a} have exhibited a striking anisotropy
for Landau level (LL) filling factors close to $\nu = N + 1/2$, with
$N\ge 4$.  A natural explanation of this anisotropy is the development
of a local 1d charge density wave (CDW) ``stripe'' state which
exhibits {\em spontaneous} orientational order, that can be further
pinned by weak crystal- or in-plane magnetic fields \cite{pan99}.
Hartree-Fock studies had predicted the existence of inhomogeneous
states \cite{fukuyama79,fogler96,moessner96}, with corroboration
coming from exact diagonalization \cite{rezayi99} and DMRG
\cite{shibata01} studies of small systems.  Such inhomogeneous states
of matter are ubiquitous in systems where there is a competition
between a repulsive, long-range (Coulomb) interaction and an
attractive, short-range (exchange) interaction.

While structurally similar to conventional CDWs in metals, in
semiconductors the CDW is expected to be only weakly pinned to the
underlying crystal because of the large disparity between the
electronic and ionic densities.  Therefore, in semiconductors a good
starting point for the CDW state is the ``quantum Hall smectic'' (QHS)
\cite{fradkin99,macdonald00,wexler01,lopatnikova01,barci01}, in which
translational and orientational orders are broken {\em spontaneously},
with crystalline fields acting as small perturbations.

Although the QHS is an interesting state in its own right, an
orientationally-ordered quantum Hall nematic (QHN)
\cite{fradkin99,balents95} liquid is sufficient to account for the
observed transport anisotropies.  Additional motivation for studying
the QHN is the instability of a 2d smectic to arbitrarily weak thermal
fluctuations \cite{toner82}, which restore translational order and
asymptotically reduce it to a nematic.  Also, at $T=0$ sufficiently
strong quantum fluctuations can unbind dislocations, driving quantum
melting of the QHS into the QHN \cite{twosmectics} and providing a
continuous transition from the QHS to the isotropic, compressible
($\nu = N+1/2$) electronic liquid \cite{halperin93}.

A conceptually useful picture of the QHN is a QHS in the presence
of a plasma of {\em free} dislocations of density $\xi_d^{-2}$. In
conventional liquid crystals, where dislocations obey simple diffusive
dynamics, this picture allows for a reliable derivation of the nematic
hydrodynamics from that of the smectic \cite{toner82,ostlund82}.
However, the $T=0$ quantum dynamics of dislocations in the QHS is
complex and is yet to be understood, as it depends upon the
dislocation's charge, statistics, and the nature of the QHS state
\cite{twosmectics}; these subtleties are not addressed in treatments
which use formal duality arguments \cite{fogler01}
or a Lorentz-invariant extension of the static model \cite{balents95}.

In this Letter we present a theory of the compressible quantum Hall
nematic, which is guided by the chiral edge dynamics
\cite{macdonald00,lopatnikova01,barci01} of the local smectic
layers. The result is a novel quantum rotor model with a ``soft''
kinetic energy, stemming from the compressibility of the QHN and the
underlying non-inertial LL dynamics.  Further support for the validity
of our model comes from a complementary analysis of the finite $T$
hydrodynamics of a charged nematic in a magnetic field, which in the
collisionless limit reproduces the microscopic dynamics.

In marked contrast to conventional liquid crystals where the
constituent molecules are highly anisotropic, in the QHN it
is the interaction-driven, dislocation-free mesoscopic regions of size
$\xi_d$ that play the role of ``nematogens,'' with each region
characterized by a local unit director field ${\bf n} ({\bf r})$
(normal to the local smectic layers).  
At scales longer than
$\xi_d$ (which at $T=1$~K we estimate to be of order 2000~\AA
\cite{wexler01}) the QHN Hamiltonian for the orientational degrees of
freedom has the standard form
\begin{eqnarray} 
 H_N& =& {1\over 2} \int d^2r\, \Big\{ \chi^{-1}(\delta \hat{\rho})^2 
\nonumber \\
& & \ + K_1 (\nab\cdot\nhat)^2 
+ K_3 [\nhat \times (\nab\times \nhat)]^2 - ({\bf h}\cdot\nhat)^2\Big\},
\label{hamiltonian}  
\end{eqnarray}
where $K_1$ and $K_3$ are the Frank elastic constants \cite{wexler01},
${\bf h}$ is a symmetry-breaking field (e.g., the crystalline
anisotropy or an in-plane magnetic field), $\chi$ is the
compressibility of this metallic state (including Coulomb interactions
replaces $\chi^{-1}$ by $\chi^{-1}+2\pi e^2/q$), $\delta
\hat{\rho}({\bf r})$ is the local electronic density {\em operator},
and in a single LL we have justifiably ignored the kinetic energy.

The quantum dynamics of the problem is encoded in the commutation
relation for the operators $\delta \hat{\rho}$ and $\nhat$. This can
be obtained by quantizing the classical dynamics of the edges of the
local smectic layers (taken parallel to the $x$-axis).  A fluctuation
in the electron density $\delta \rho$ leads to a force
$-\partial_x(\chi^{-1}\delta\rho)$ parallel to the layers, which in
the high field limit is balanced by the Lorentz force $-e B \dot{u}$;
a {\em uniform} compression along the stripe leads to a layer
displacement $u\hat{\bf y}$, consistent with the fact that the
LL-projected particle coordinates $X$ and $Y$ do not commute
(equivalently encoded in the dynamics of chiral edge bosons
\cite{macdonald00,lopatnikova01,barci01}). Consequently, a {\em
  nonuniform} compression along the stripe rotates the stripe through
an angle $\theta\equiv\partial_x u$ determined by
$eB\dot{\theta}=\partial_x^2(\chi^{-1}\delta\rho)$.  Quantizing this
dynamics on scales longer than $\xi_d$ \cite{USunpublished} leads to
the commutation relation
\begin{equation}
[ \delta \hat{\rho}({\bf r}), \hat{\bf n} ({\bf r}')] = i\ell^2
\hat{\bf z}\times\hat{\bf n}({\bf r}') \partial_x^2 \delta({\bf r-r'}),
\label{commutation}
\end{equation}
where $\ell=\sqrt{1/eB}$ is the magnetic length and in our units
$\hbar=k_B=1$.  The corresponding QHN action 
$S_N= \int dt\, d^2r {\cal L}_N$ is specified by a novel $O(2)$ 
``soft'' quantum rotor model, with a Lagrangian density
\begin{eqnarray} 
{\cal L}_N &=& (eB) {\bf L}\cdot(\hat{\bf n}\times\partial_t\hat{\bf n})
  - {1\over 2}\Big\{\chi^{-1} (\partial_x^2 {\bf L})^2 
\nonumber \\
& & + K_1 (\nab\cdot\hat{\bf n})^2 
+ K_3 \left[\hat{\bf n}\times(\nab\times\hat{\bf n})\right]^2 
- ({\bf h}\cdot\hat{\bf n})^2 \Big\},  
\label{lagrangian} 
\end{eqnarray}
where ${\bf L} = \hat{L}_z \hat{\bf z}$ is the effective angular
momentum operator conjugate to $\hat{\bf n}$, with commutation
relation $[\hat{L}_z({\bf r}),\hat{\bf n}({\bf r}')] = i \ell^2
\hat{\bf z}\times\hat{\bf n}({\bf r}')\delta({\bf r-r'})$ and defined
by $\delta\hat{\rho} = \partial_x^2\hat{L}_z$. In addition to the
first ``Berry's phase'' term associated with the conserved
$z$-component of the angular momentum, a notable feature of the QHN is
that the kinetic energy (the second term) vanishes at long
wavelengths, a property which can be traced to the non-inertial 2d
electron dynamics in a strong magnetic field.

Having defined the model, we can now assess the effect of fluctuations
on the QHN. Parameterizing the director as ${\bf
n}=(\sin\theta,\cos\theta)$, we compute the (imaginary) time-ordered
orientational correlation function $C({\bf
r},\tau)=\langle{\cal{T}}_\tau\theta({\bf r},\tau)\theta({\bf
0},0)\rangle$, whose Fourier transform $\tilde{C}({\bf q},\omega_n)$
at Matsubara frequency $\omega_n=2\pi n T$ can be measured in
depolarized dynamic light scattering experiments.  We find
\begin{equation}
\tilde{C}({\bf q},\omega_n) 
= T\ell^4 \chi^{-1} {q_x^4 \over \omega_n^2 + \epsilon_{\bf q}^2},
\end{equation}
with an anisotropic dispersion $\epsilon_{\bf q}$ for orientational
fluctuations about a $y$-directed nematic state given by
\begin{equation}
\epsilon_{\bf q}= \pm \ell^2 \chi^{-1/2} q_x^2
\sqrt{ K_1 q_x^2 + K_3 q_y^2 + h^2},
\label{goldstone}
\end{equation}
exhibiting a line of nodes along the nematic order. In contrast to
naive expectations based upon experience with conventional nematics,
in the QHN, because of the strong magnetic field, this orientational
mode remains {\em gapless} ($\epsilon_{\bf q}\sim q_x^2$) even in the
presence of an ordering field ${\bf h}=h\hat{\bf y}$.  For $h=0$ and
in a single elastic constant approximation ($K_1=K_3=K$) the spectrum
is $\epsilon_{\bf q}\propto q_x^2 |q|$ ($\propto q_x^2|q|^{1/2}$ with
Coulomb interactions).

Standard analysis shows that at $T=0$ the equal-time connected
correlation function $C_c({\bf r},0)\equiv C({\bf 0},0)-C({\bf r},0)$
saturates at $C_c({\bf \infty},0)\equiv\theta_{rms}^2\sim
c_1\ell^2/(\xi_d^3\sqrt{K\chi})$ ($c_1=O(1)$ constant), and therefore
predicts that a 2d QHN exhibits true long-range order at $T=0$, with
the nematic order parameter $\psi_2=\langle e^{2i\theta}\rangle$
reduced by quantum fluctuations from its $T=0$ classical value of $1$
to $\psi_2(0,h)\approx e^{-2\theta_{rms}^2}<1$.  At low $T$, such that
$T\ll T_Q\equiv \ell^{2}\xi_d^{-3}\sqrt{K/\chi}$, $C_c({\bf r},0)$
exhibits a plateau for
$\xi_d<r<\xi_T\equiv\left(\ell^2/T\right)^{1/3}(K/\chi)^{1/6}$, but
(for $h=0$) asymptotically crosses over to classical logarithmic
growth $\sim\ln({r/\xi_T})$ characteristic of quasi-long-range order
of 2d classical nematics. For $h\neq0$ the order is long-ranged even
at finite $T$, but is reduced from the quantum-renormalized $T=0$
value of $\psi_2(0,h)$ down to $\psi_2(T,h)= \psi_2(0,h)R(T,h)$, with
$R(T,h) = \exp{[-S\left({\xi_T^3}/{\xi_{h0}^3}\right)
  \ell^2/(8\pi^2\xi_{h0}^3\sqrt{K\chi})]}$, $S(y)=\int_0^{\Lambda^2}
dz z(z+1)^{-1/2}\int_0^{2\pi}
d\theta\cos^2\theta\left[\coth(\frac{y}{2}z\sqrt{z+1}\cos^2\theta)-1\right]$,
$\Lambda=\xi_{h0}/\xi_d$, and $\xi_{h0}=K^{1/2}/h$ a strong-coupling
orientational pinning length associated with $h$.  In the quantum
regime, $T\ll T_h\equiv h^3\ell^2/(\chi^{1/2}K)<T_Q$ (equivalently,
$\xi_d<\xi_{h0}\ll\xi_T$), we find $S(y)\approx c_2/y^{3/2}$
($c_2\approx \zeta(3/2)/(3 \pi^{3/2})\approx 0.16$), leading to
$R(T,h)\approx e^{-(T/T_0)^{3/2}}$, with $T_0\equiv \ell^{2/3}K^{1/3}
h /(c_2^{2/3}\chi^{1/6})$. Such nonanalytic (in $T$) thermal
suppression of nematic order contrasts strongly with the naive
classical spin-wave prediction of {\em linear} reduction with $T$
\cite{fradkin99}. In the opposite, semi-classical $T_h\ll T < T_Q$
regime (equivalently, $\xi_d<\xi_T\ll\xi_{h0}$),
$R(T,h)=f(\xi_T/\xi_h)$, with $f(y)\approx y^{\eta_2/2}$,
$\eta_2=2 T/\pi K$, and $\xi_h=\xi_T(\xi_h^0/\xi_T)^{4/(4-\eta_2)}$
the semi-classical pinning length, strongly thermally
renormalized upward from its strong-coupling value $\xi_{h0}$.  Even
in this semi-classical regime, quantum effects are observable with $T$
entering through {\em both} $\eta_2(T)$ and the quantum UV cutoff
$\xi_T(T)$, only crossing over to the conventional classical result
for $T>T_Q$, corresponding to $\xi_T(T)<\xi_d$.

Next, we turn to the effects of quenched disorder on the QHN, with the
dominant effects at long length scales coming from the random
anisotropy field $\delta{\bf h}({\bf r})$ arising from, e.g., crystal
defects and sample roughness, both randomly pinning the nematic
director.  Since the disorder is static, in a linear ``Larkin''
approximation (valid at scales smaller than disorder persistence
length $\xi_\Delta$) the quantum dynamics drops out and we can adopt
many recent results on the effects of disorder on classical liquid
crystals \cite{radzihovsky99}.  For $h=0$ we find that for $d<4$ the
nematic state is unstable to infinitesimally weak quenched disorder;
in 2d the local orientational order only persists out to the Larkin
length $\xi_L\approx(2\pi)^{3/2}K/\Delta^{1/2}$, where $\Delta$ is the
variance of the random orientational field $\delta h_x({\bf r})\delta
h_y({\bf r})$.  On longer scales the full nonlinearity of the random
field must be taken into account; a preliminary analysis suggests the
existence of a quantum nematic glass \cite{USunpublished} akin to its
classical analog in nematic liquid crystals \cite{radzihovsky99}.

Tunneling can provide important spectroscopic information about the
low-energy excitations.  In order to calculate the tunneling current
we need to construct an electron creation operator $\psihat^\dagger$,
satisfying $[\delta\hat{\rho}({\bf r}),\psihat^\dagger({\bf r}')] =
\psihat^\dagger({\bf r}')\delta ({\bf r-r}')$; the operator
\begin{equation} 
\psihat^\dagger({\bf r},t) = {1\over \sqrt{2\pi\ell^2}}\, e^{-i\int_{\bf r'}
\,\left[D({\bf r}-{\bf r}')\hat{\theta}({\bf r}',t) +
Arg({\bf r}-{\bf r}')\ell^2\partial_x^2\hat{L}_z({\bf r}',t)\right]},
\end{equation} 
accomplishes this, with $D({\bf r}-{\bf r}')$ a solution of
$\ell^{2}\partial_x^2 D({\bf r}-{\bf r}') = \delta({\bf r}-{\bf r}')$
and a vortex attachment factor ensuring the fermionic statistics.  The
tunneling I-V is determined by the Laplace-transform of the local
imaginary-time Green's function $G(\tau) = \langle {\cal T}_\tau
\psihat({\bf r},0)\psihat^\dagger ({\bf r}, \tau)\rangle \propto
e^{-\Phi(\tau)}$, with $\Phi(\tau) = \frac{\chi^{-1}}{2}\int\frac{d^2
q}{ (2\pi)^2}(1 - e^{-\epsilon_{\bf q} |\tau|})/\epsilon_{\bf q}$. For
short-range interactions, the leading behavior is 
$\Phi(\tau)\approx\left(|\tau|/\tau_0\right)^{1/2}$
($\tau_0\equiv\frac{1}{2}\pi^3 K^{1/2}\chi^{3/2}\hbar\ell^2\xi_d$), 
independent of the $\xi_\tau/\xi_h$ ratio
($\xi_\tau^3=\tau\ell^2\sqrt{K\chi^{-1}}/\hbar$). Tunneling into the QHN
is therefore suppressed relative to both the isotropic $\nu = 1/2$
Halperin-Lee-Read (HLR) state \cite{tunnelingCF} and the QHS
\cite{lopatnikova01,barci01}. Coulomb interactions modify the large
$h$ limit, giving $\Phi(\tau)\sim |\tau|^{3/5}$, but leave the small
$h$ regime unchanged.

Although above model of the QHN is quite appealing and {\em
mathematically} self-contained, on general physical grounds a theory
of a {\em compressible} QHN, based on a collective nematic
orientational degree of freedom $\theta({\bf r},t)$ {\em alone} is
necessarily incomplete. This shortcoming is clear from the defining
properties of the QHN state, a {\em compressible} anisotropic {\em
liquid}, in which free dislocations and their bound states of
vacancies and interstitials (bubbles) constitute additional low-energy
quasi-particle degrees of freedom. This contrasts strongly with the QHS,
where a $T=0$ description in terms of collective phonon (chiral edge
boson) degrees of freedom alone is physically consistent
\cite{fradkin99,macdonald00,wexler01,lopatnikova01,barci01}, although
not the only possibility \cite{twosmectics}. 

Consequently, we must extend above QHN model to include these
additional quasi-particle degrees of freedom.  Having been
unable to do so from first principles, we conjecture that the
quasi-particle sector of the full model is described by the HLR
Lagrangian, ${\cal L}_{\rm HLR}[\hat{\psi},{\bf a}]$, with
$\hat{\psi}({\bf r})$ and ${\bf a}({\bf r})$ the
composite fermion (CF) and statistical gauge fields \cite{halperin93},
coupled to the nematic orientation collective degree of freedom
$\hat{\bf n}({\bf r})$ via
\begin{equation} 
{\cal L}_{\rm int} =g_m Q_{ij} D_i^* \hat{\psi}^\dagger  D_j \hat{\psi}
+(g_s|{\bbox\nabla}\cdot{\hat{\bf n}}|^2
+ g_b|{\bbox\nabla}\times\hat{\bf n}|^2)\hat{\psi}^\dagger\hat{\psi}.
\end{equation} 
This coupling is dictated by symmetry, with the first term the
anisotropic CF mass proportional to the nematic order parameter
$Q_{ij}= n_i n_j - (1/2)\delta_{ij}$ [with $\psi_2=2(Q_{xx}+i
Q_{xy})$], the second and third terms the coupling of the CF density to
splay and bend, and ${\bf D}={\bbox\nabla}-i{\bf a}$.

The success of the HLR state at $\nu=1/2$ \cite{halperin93} suggests
that it might also be a good starting point at higher partially filled
LLs, $\nu\approx N + 1/2$.  Spontaneous development of
orientational order (facilitated by screening of the Coulomb
interaction by lower LLs) then leads to an orientationally-ordered
HLR state. Such a state naturally coincides with the above picture of
the compressible QHN as a melted QHS. The relation
$\partial_x\rho_{i-v}-\partial_y\theta=b_d({\bf r})$ between the
interstitial-vacancy density, $\rho_{i-v}=\rho_{i}-\rho_{v}$ and
dislocation density $b_d({\bf r})$ is reminiscent of the
relation ${\bbox{\nabla}}\times {\bf a}=\rho_{cf}$ \cite{halperin93}.
This intriguingly suggests the identification of the gauge field
with vacancies and interstitials [i.e., ${\bf a}({\bf
  r})=(\theta,\rho_{i-v})$], and 
CF fermions with dislocations, lending
further support to the above conjecture.

The presence of these HLR quasi-particles will undoubtedly lead to
Landau-like damping of the nematic director dynamics. Nematic order
will in turn induce a spontaneous Fermi surface
distortion \cite{oganesyan01} and the concomitant highly anisotropic
transport observed in experiments \cite{lilly99a,pan99}.  The
associated director $\hat{\bf n}$ fluctuations will also contribute to
the CF self-energies, modifying them relative to those of the isotropic
HLR state.  The quasi-particle sector of the model will allow for a
second tunneling channel, which, to lowest order, should add in
parallel 
to the tunneling results discussed above.  
We leave the analysis of these and other related
problems to a future publication \cite{USunpublished}. 

We now turn to hydrodynamics, which 
provides further support for our microscopic model of the QHN. Focusing on the
Goldstone director mode $\hat{\bf n}$ and conserved particle ($\rho$)
and momentum ($\rho{\bf v}$) densities (ignoring for simplicity the energy
density), standard methods lead to a set of hydrodynamic equations
for an orientationally-ordered 2d liquid of charged rods in the
presence of a magnetic field ${\bf B}=B\hat{\bf z}$, which to linear
order are
\begin{eqnarray}
&&\partial_t \rho= - \rho_0 \nab\cdot {\bf v},\\
\label{hydro1}
&&\partial_t v_x =- {1\over 2m \rho_0}(1-\lambda)\partial_y
\left({\delta  H_N\over \delta \theta}\right)
-  \partial_x\left({\delta  H_N \over \delta \rho}\right)
 + \omega_c v_y \nonumber\\
&&-\gamma_{xx} v_x -\gamma_{xy} v_y 
+ \nu_{11}\partial_x^2v_x + \nu_{33}\partial_y^2 v_x
 + (\nu_{12}+ \nu_{33})\partial_x\partial_y v_y,\nonumber\\
\label{hydro2}
&&\partial_t v_y= {1\over 2m \rho_0}(1+\lambda)\partial_x
\left({\delta  H_N\over \delta \theta}\right)
-  \partial_y \left({\delta  H_N \over \delta \rho}\right)
-\omega_c v_x \nonumber\\ 
&&-\gamma_{yx} v_x -\gamma_{yy} v_y
+ \nu_{22}\partial_y^2v_y + \nu_{33}\partial_x^2 v_y
 + (\nu_{12}+ \nu_{33})\partial_x\partial_y v_x,\nonumber\\
\label{hydro3}
&&\partial_t\theta = {1+\lambda\over 2} \partial_x v_y
 - {1-\lambda \over 2} \partial_y v_x
-\gamma_\theta {\delta  H_N \over \delta \theta},\nonumber
\label{hydro4}   
\end{eqnarray}
where $H_N$ is given by Eq.\ (\ref{hamiltonian}), $\omega_c\equiv
eB/m$, $\nu_{ij}$'s are viscosities, and $\gamma_\theta$ and
$\gamma_{ij}=\gamma_0\delta_{ij}+ \gamma_Q Q_{ij}$ are
the orientational and translational (momentum) relaxation rates. The
{\it reversible} coupling $\lambda$ (determined by the
microscopic physics \cite{forster74}, and generically expected\cite{ostlund82}
to approach $1^-$ near the QHN-QHS transition), encodes the
competition between the director's rotational convection and shear
alignment.  As in conventional nematics, for $|\lambda|\ge 1$ the
director will shear-align at an angle $\phi_\lambda$ to a prescribed
uniform shear flow ${\bf v}=y\hat{\bf x}$, with $\tan\phi_\lambda =
\sqrt{(\lambda + 1)/(\lambda -1)}$; for $|\lambda|<1$, such a steady
state is impossible and $\hat{\bf n}$ will tumble anisotropically (via
instantons).

Analysis of Eqs. (\ref{hydro1}) in the collisionless and
Galilean-invariant limit ($\nu_{ij}=\gamma_\theta = \gamma_{ij}=0$)
leads to four modes, only two of which (corresponding to a mixture of
density $\rho$ and nematic orientation $\theta$) are hydrodynamic and
which propagate with a dispersion
\begin{equation}
\omega_{\bf q} ={\pm\ell^2 \over 2\chi^{1/2}}
\left[ (1+\lambda) q_x^2
+ (1-\lambda)  q_y^2 \right] 
\sqrt{ K_1 q_x^2 + K_3 q_y^2 + h^2}. 
\label{nematic10}
\end{equation}
Analysis of the eigenmodes reveals that this novel $q^3$ mode
corresponds to clockwise/counter-clockwise director oscillations
beating against density fluctuations.  For $\lambda = 1$ this mode is
{\it identical} to the microscopic, $T=0$ dispersion in Eq.\ 
(\ref{goldstone}).  This, together with the observation that the long
time limit of hydrodynamic equations can be obtained from a Poisson
bracket between $\rho$ and $\theta$, generalized to an arbitrary
$\lambda$ \cite{USunpublished}, suggests that the microscopic
commutation relation, Eq. (\ref{commutation}), 
must be similarly generalized to
\begin{equation}
\hspace{-0.3cm}
[ \delta \hat{\rho}({\bf r}), \hat{\bf n} ({\bf r}')]={i\over 2} \ell^2
 \hat{\bf z}\times \hat{\bf n}({\bf r}')
\left[ (1+\lambda) \partial_x^2  
+ (1-\lambda)\partial_y^2\right] \delta({\bf r-r'}),
\label{commutation2}
\end{equation}
with a corresponding replacement in the Lagrangian
(\ref{lagrangian}) of $\partial_x^2{\bf L} \rightarrow
\frac{1}{2}\left[(1+\lambda)\partial_x^2 +
  (1-\lambda)\partial_y^2\right]{\bf L}$.
A notable feature of the dispersion (\ref{nematic10}) 
in the alignable $|\lambda|>1$
regime are lines of nodes along $q_y/q_x = \pm\tan\phi_\lambda$, a
property that should reflect itself in a variety of experiments.  For
instance, our tunneling result for $\lambda=1$ and short range interactions,
$\Phi(\tau)\sim |\tau|^{1/2}$, will become $\sim|\tau|^{1/3}$
for $h=0$ and $\lambda\neq 1$, and for $h\neq0$ will become
$\sim\log\tau$ ($|\lambda|<1$) and $\sim\log^2\tau$ ($|\lambda|>1$).

In the presence of dissipation, at sufficiently long scales the two
density-director modes become overdamped. For the Galilean-invariant
case ($\gamma_{ij}=0$) the relaxation rates are $\omega_-=-i D_- q^4$
and $\omega_+ = - i D_+ (K_1 q_x^2 + K_3 q_y^2)$, with $D_\pm$
model dependent diffusion coefficients. However, because
QHN exhibits finite conductivity, at sufficiently long scales momentum
relaxation becomes important and the density and director dynamics
decouple, leading to diffusive relaxation rates $\omega_\rho=i D_\rho
q^2$ and $\omega_\theta=i D_\theta (K_1 q_x^2 + K_3 q_y^2)$.

Although the velocity modes are non-hydrodynamic in the presence of a 
magnetic field, their relaxation determines the linear
longitudinal transport.  In the isotropic, $\psi_2=0$ (HLR) state,
this leads to the Drude model with an isotropic resistivity.  In the QHN,
$\psi_2\neq0$ and we find a resistivity anisotropy
$(\rho_{xx}-\rho_{yy})/(\rho_{xx}+\rho_{yy})=\psi_2\gamma_Q/\gamma_0$
proportional to the nematic order parameter \cite{fradkin99}, 
making the $h$ and $T$ dependence of the order parameter 
experimentally accessible. 
 
In summary, by combining detailed microscopic and 
hydrodynamic analyses, we have proposed a model for a
QHN, and argued that this novel orientationally-ordered electronic
liquid crystal state may be realized in nearly half-filled high LLs.
We predict the existence of a novel director-density mode, with
a highly anisotropic $q^3$ dispersion, which remains gapless even in the
presence of an ordering field.  The nematic order is
long-ranged at $T=0$, but is destroyed by thermal fluctuations and
quenched disorder.  Clearly,
many interesting theoretical and experimental questions remain and now
can be addressed in detail using the model presented
here \cite{USunpublished}. 

We thank M.P.A. Fisher, M. Fogler, S. Girvin, B. Halperin and C.
Wexler for comments, and acknowledge support from NSF DMR-9978547
(ATD), DMR-9625111 (LR), and the Sloan and Packard Foundations (LR).
LR thanks Harvard's Physics Department for its hospitality.
 
\vspace{-1.0\baselineskip}

\end{document}